# Smaller, Younger, and More Visible? AI-Assisted Writing, Scientific Team Composition, and the Visibility of Scientific Knowledge

Mingze Zhang[1], Haoyang Wang[2], Yi Bu[3], Star Xing Zhao[4], Meijun Liu[5*]

[1] Department of Data Science, City University of Hong Kong, Hong Kong SAR, China
[2] School of Information, University of Texas at Austin, Austin, TX 78701, USA
[3] Department of Information Management, Peking University, Beijing 100871, China
[4] Institute of Big Data, Fudan University, Shanghai 200433, China
[5] Institute for Global Public Policy, Fudan University, Shanghai 200433, China

* Corresponding author: meijunliu@fudan.edu.cn

## Abstract

Generative AI is reshaping scholarly communication, yet little is known about how AI-assisted writing relates to the organization and visibility of scientific knowledge production. This study examines the associations between AI-modified text, publication-team characteristics, and scholarly visibility. Drawing on 144,930 full-text articles and reviews from the PLoS family and the Nature Portfolio published between 2020 and 2025, we use a novel text-based detection approach to estimate AI usage scores as indirect indicators of AI-assisted writing. This indicator does not directly capture actual AI use in academic writing but provides an indirect signal of AI-assisted writing. We focus on two dimensions of scholarly communication, the production and dissemination of scientific knowledge. The former is reflected by two team characteristics, and the latter is measured by short-term citations. We examine AI usage score's associations with team academic age, team size, and citations received within 365 days of publication. Using descriptive comparisons, multivariate OLS, quantile, and Poisson regressions, and propensity score matching, we find that higher AI usage scores are associated with lower average team academic age and smaller publication teams, although the statistical evidence of the estimated negative association between AI usage score and team size is weaker in the Nature Portfolio. Higher AI usage scores are also positively associated with one-year citation counts. We further find that higher AI usage score is related to a weaker visibility advantage for larger teams. These findings suggest that AI-associated writing signals are associated with differences in publication-team composition and greater short-term scholarly visibility. The study contributes to understanding how generative AI relates to the evolving processes of scholarly communication, with implications for research evaluation, funding, and academic training.

## 1 Introduction

Scientific knowledge production has increasingly relied on larger, more collaborative, and geographically dispersed teams (Hsiehchen et al., 2015; Liu, Bu, et al., 2022; Wuchty et al., 2007). This trend has often been attributed to the growing "burden of knowledge," which requires greater specialization and coordination among researchers (Jones, 2009). The rapid development of generative artificial intelligence (GenAI), raises new questions about these established patterns. GenAI can assist with tasks traditionally performed by researchers (Qin et al., 2026), particularly academic writing (Harries et al., 2025), and may change the resources required to produce and communicate scientific knowledge (Binz et al., 2025; Kusumegi et al., 2025; Wang et al., 2023).

Scholarly communication encompasses the production, evaluation, dissemination, and subsequent uptake of scientific knowledge (Borgman, 2007). Within this process, publication teams represent the organizational context of knowledge production, while citations provide observable indicators of the subsequent visibility (Tomaszewski, 2023). Examining AI-assisted writing in relation to both team composition and citation visibility therefore offers a way to understand how GenAI is associated with changes across two interconnected dimensions of scholarly communication.

We first examine whether AI-assisted writing is associated with differences in the composition of scientific teams. Although team composition encompasses multiple dimensions, such as diversity (Nielsen et al., 2017; Yang et al., 2022), hierarchical structure (Xu et al., 2022), we focus on team size and academic age. This is because they capture two fundamental dimensions of human resources in scientific knowledge production: collaborative labor and accumulated academic experience. GenAI may alter these requirements by assisting with tasks previously performed by collaborators and by providing research support that may be particularly valuable to less-experienced researchers. Team size, measured by the number of co-authors (Larivière et al., 2015; Thelwall & Maflahi, 2022), captures the scale of authorship-based collaboration, whereas team academic age, measured by the average academic age of team members (Xu et al., 2022; Yang et al., 2024), serves as a proxy for the team's accumulated academic experience.

The relationship between AI-assisted writing and scientific team composition may follow two competing mechanisms. The substitution mechanism suggests that GenAI can partially replace certain forms of human assistance in scientific writing, reducing the resources required for manuscript preparation. Such benefits may be particularly valuable to smaller teams, early-career researchers, and non-native English speakers who face greater resource or linguistic constraints (Andersen et al., 2025; Ben-Zion et al., 2026; Dorta-González et al., 2024). However, the complementarity mechanism suggests that effective AI use depends on existing research expertise, institutional resources, and specialized capabilities (Bianchini et al., 2025; Hey, 2023; Hosseinioun et al., 2026). Larger and more experienced teams may be better positioned to integrate GenAI into their research workflows (Bianchini et al., 2025). These competing perspectives suggest that AI-assisted writing may be associated with different patterns of team size and academic age, making the direction of these relationships uncertain.

A further question concerns the dissemination of scientific knowledge, whether AI-assisted writing translates into greater scientific visibility, as reflected in citations. Citations provide observable traces of formal scholarly connections between publications and are commonly used as indicators of scholarly visibility and recognition (Taşkın, 2026; Thelwall et al., 2023). Clear and well-crafted academic writing can make research findings more accessible, and potentially broaden its readership (Ryba et al., 2019). On the one hand, AI-assisted writing may improve the clarity, cohesion, readability, and accessibility of scientific publications (Ibrahim & Mahmoud, 2026), making research findings easier to communicate and potentially attracting more citations. On the other hand, improved scientific communication does not necessarily translate into greater scholarly recognition. Citations depend not only on how clearly research is presented but also on its novelty, substantive scientific value, and relevance to ongoing scientific debates (Aksnes et al., 2019). Thus, although AI-assisted writing may facilitate the communication and dissemination of scientific knowledge, its relationship with citations remains unclear.

Previous research has shown that team size and academic experience are important characteristics of scientific collaboration, often associated with differences in research performance and scientific visibility (Larivière et al., 2015; Wuchty et al., 2007; Xu et al., 2022). Larger and more experienced teams may benefit from greater resources, accumulated expertise, and broader collaboration networks (Wu et al., 2019). However, whether these advantages persist in the context of AI-assisted scientific writing remains unclear. AI writing tools may provide additional support for researchers in preparing and communicating their work, potentially changing how team size and academic experience are associated with scientific visibility.

Despite growing interest in GenAI, existing research has mainly focused on individual researchers' adoption of AI and its effects on writing practices and productivity (Kusumegi et al., 2025; Ma et al., 2026; Zhao & Li, 2026). Less is known about how AI-assisted writing relates to the composition of scientific publication teams and the subsequent visibility of their research.

To address these questions, we analyze 144,930 full-text publications from the PLoS family and the Nature Portfolio published between 2020 and 2025. We estimate the extent of AI-modified text using a novel text-based detection approach (Liang et al., 2025; Qian et al., 2026) and link these estimates to publication and author metadata. AI-assisted writing is inferred from publication text rather than directly observed, and we treat the resulting AI usage score as an indirect measure of AI-assisted writing and focus on its associations with team characteristics and short-term citations. Specifically, we address four research questions:

RQ1: How is AI usage score associated with publication team academic age?

RQ2: How is AI usage score associated with publication team size?

RQ3: How is AI usage score associated with the short-term visibility of scientific knowledge, as reflected in citation counts?

RQ4: Does AI usage score moderate the relationships between publication-team size, team academic age, and the short-term citation visibility of scientific knowledge?

Our study makes three contributions. First, it contributes to the literature on scholarly communication and the organization of scientific research by examining how AI-assisted writing is associated with research team composition and scientific visibility. By considering both team size and academic age, it extends existing discussions of whether GenAI may lower barriers to scientific communication or reinforce advantages associated with established research resources and capabilities. Furthermore, by examining how AI usage moderates the relationships between team characteristics and citation visibility, the study provides new insights into the potential role of GenAI in shaping existing inequalities in scientific recognition.

Second, the study provides publication-level empirical evidence on AI-assisted scientific writing, complementing previous research largely based on researchers' self-reported adoption and perceptions of GenAI. By examining AI usage scores alongside team size, academic age, and subsequent citation counts across two major publication portfolios, it offers a systematic assessment of how AI-assisted writing signals are associated with scientific knowledge production and dissemination.

Third, the findings have practical and policy implications for researchers, research institutions, and science policymakers. They inform discussions about how AI writing tools may support scientific communication across teams with different resources and levels of experience.

# 2 Related Work

## 2.1 Generative AI and Scientific Teamwork

Scientific research has become increasingly collaborative as the growing complexity of knowledge requires greater specialization and division of labor. Earlier technological advances facilitated collaboration by reducing communication and coordination costs across institutional and geographical boundaries (Ding et al., 2010). GenAI may reshape scientific teamwork in a different way. Rather than simply facilitating collaboration, LLMs can assist with literature synthesis, programming, manuscript preparation, and language editing (Dwivedi et al., 2023; Kusumegi et al., 2025; Wu & Vasilescu, 2026). These capabilities may reduce the need for certain forms of human efforts in scientific work. Recent studies have reported reductions in team size among non-native English-speaking authors following the release of ChatGPT (Ben-Zion et al., 2026), a rebound in single-authored publications across several disciplines (Matsui, 2026), and changes in the division of labor within scientific teams (Zheng et al., 2026). Although these studies do not establish causal effects of GenAI, they suggest that the organization of scientific work may be changing.

Early-career researchers may benefit more from these developments. They often have less accumulated experience, smaller professional networks, and fewer institutional resources, while facing considerable pressure to publish (Liu et al., 2023; Nicholas et al., 2024). These barriers may increase their motivations to apply GenAI in their work. GenAI may help address some of these constraints by supporting research tasks and academic writing. Yet access to AI tools does not necessarily translate into equal benefits. Researchers' scientific capabilities, collaborative networks, and institutional resources may influence their ability to adopt and use AI effectively (Bianchini et al., 2025). Recent evidence reported AI-related resource reallocation toward larger research teams (Hosseinioun et al., 2026).

These findings point to two competing mechanisms. Under a substitution mechanism, GenAI reduces the need for certain human work, potentially benefiting smaller teams and less-experienced researchers. Under a complementarity mechanism, AI enhances existing research capabilities and may favor larger and more established teams. We have little knowledge of whether AI-assisted writing is associated with smaller or larger teams, and with younger or more experienced researchers. We examine these relationships through team size and team academic age.

### 2.2 Generative AI and the Visibility of Scientific Knowledge

Scientific publications are central to the dissemination and subsequent use of scientific knowledge. Citations provide observable indicators of scholarly recognition and uptake, although they are influenced by journal reputation, author characteristics, disciplinary context, and other factors beyond scientific quality (Tahamtan et al., 2016). How research is written and presented may also affect its visibility. Previous studies have linked readability and promotional language to publication outcomes, citations, and scholarly attention (Marino Fages, 2020; Stavrova et al., 2025). However, the association between writing characteristics and citations varies across research contexts (Ante, 2022; Lei & Yan, 2016).

Language barriers further shape the circulation of scientific knowledge. The dominance of English in international academic publishing places additional demands on non-native English-speaking researchers (Drubin & Kellogg, 2012). Di Bitetti and Ferreras (2017) found that articles published in languages other than English tend to receive fewer citations. GenAI may help researchers overcome some of these barriers by assisting with manuscript drafting, translation, and language editing. Its use in scholarly communication has grown rapidly (Arroyo-Machado et al., 2025; Nicholas et al., 2024), with studies documenting an increasing presence of AI-assisted writing in scientific publications (He & Bu, 2026; Liang et al., 2024).

AI-assisted writing may improve the clarity and accessibility of publications, making research findings easier to communicate and potentially increasing their visibility. However, better writing does not necessarily lead to more citations. Scholarly recognition also depends on the novelty, relevance, and substantive contributions of research (Aksnes et al., 2019; Tahamtan et al., 2016). GenAI may improve how scientific knowledge is presented without changing the knowledge itself. The relationship between AI-assisted writing and citations is therefore not straightforward.

Most existing studies of GenAI in academic writing have focused on its adoption (Gross et al., 2026; Ibrahim & Mahmoud, 2026), and linguistic characteristics (Bao et al., 2025), with less attention to the subsequent scientific visibility of AI-assisted publications. Citations also develop over time, and early recognition does not necessarily reflect long-term influence (Kosteas, 2018). We therefore examine whether AI-assisted writing is associated with short-term scholarly visibility, measured by citations accumulated within one year of publication.

## 3 Data & methods

### 3.1 Data source

The main data sources of this study include the PLoS (Public Library of Science) family, the Nature portfolio and OpenAlex. We investigate the PLoS family and the Nature Portfolio for two reasons. First, both provide large-scale full-text publication data across multiple scientific disciplines, allowing us to estimate AI-modified content beyond titles and abstracts. This represents a significant advantage in accuracy over other widely used bibliometric datasets, such as Web of Science, Scopus, and OpenAlex, which only provide publication metadata like titles and abstracts. Second, the two publishing portfolios cover different sets of journals and publishing contexts, allowing us to examine whether the associations between AI usage scores, publication team characteristics, and citations are consistent across these settings. We therefore conducted analyses using the pooled sample and separately for each publishing portfolio.

The PLoS family[1] is a nonprofit publisher of open-access, peer-reviewed journals covering scientific disciplines across life sciences, health, sustainability, engineering and technology. The Nature portfolio[2] encompass Nature journals, such as Nature, Nature Biomedical Engineering, Nature Human Behaviour and so forth. We downloaded the full-text corpus of the PLoS family publications from the repository of all PLoS XML article files in September 2025.[3] We derived the 15 journals data of Nature portfolio dataset provided by Liang et al. (2025). The PLoS family dataset includes full text of 105,994 publications published by 14 journals between January 2020 and May 2025. Nature portfolio dataset consists of 41,080 publications published from January 2020 to September 2024.

[1] https://PLoS.org/
[2] https://www.nature.com/nature-portfolio
[3] https://github.com/PLoS/allofPLoS

OpenAlex is another important data source. It is an open-access and comprehensive bibliographic database (Priem et al., 2022). It indexes over 200 million publications from at least 109,000 global institutions and about 124,000 venues. Due to its comprehensive coverage and open-access nature, it has been widely adopted in relevant research (Culbert et al., 2025). Another prominent advantage of OpenAlex lies in its advanced and automated name disambiguation for authors and institutions,[4] which facilitates calculation of mean career age of authors in each publication team. It provides diverse metadata of scientific publications, including publication year, author affiliations, discipline, citations, and geographic data. We downloaded the snapshot of this dataset from its official website in July 2026.[5] To obtain bibliographic and citations data of each publication in the two datasets, we linked each publication to the OpenAlex dataset using Digital Object Identifiers (DOIs), with all publications linked to OpenAlex.

As this study examines publication teams, we excluded single-authored publications and focused on collaborative publications. We further restricted the sample to English-language publications classified as either "article" or "review" in OpenAlex. After applying these criteria, the final sample consisted of 103,967 publications from the PLoS family and 40,963 from the Nature Portfolio, all published since January 2020. Of these, 48,968 PLoS publications and 14,583 Nature Portfolio publications were published after November 30, 2022, following the public release of ChatGPT.

## 3.2 Variables

The unit of analysis is the publication, with the author team listed on each publication serving as the unit for analysing team characteristics. Following established practice (Leahey, 2016; Liu et al., 2022; Wu et al., 2019), we use the term, publication team, to refer to the group of authors listed on a research article. Although research teams are generally understood as groups of individuals who work together toward shared goals (Guzzo & Dickson, 1996), the author list does not necessarily capture all individuals involved in a research project. Our analysis therefore focuses on publication team characteristics rather than the full composition or internal organization of research teams.

The main explanatory variable is AI usage score, which estimates the extent of AI-modified content in the full text of each publication. The outcome variables capture three dimensions: team academic age, publication team size, and short-term scientific visibility, measured by the number of citations received within one year (365 days since the date of publication).

[4] https://help.openalex.org/hc/en-us/articles/24347048891543-Author-disambiguation
[5] https://openalex.org/

#### *3.2.1 AI usage score*

AI-assisted writing refers to the use of AI tools to support the preparation and revision of academic manuscripts. Because the actual use of these tools is usually not directly observable, we used a text-based measure to estimate the extent of AI-modified content. This approach has been widely applied in the research on how AI contributes to academic writing (Gross et al., 2026; Liang et al., 2025; Qian et al., 2026).

Following Liang et al. (2025), we calculate an AI usage score for each publication based on its full text. This measure builds on the distributional GPT quantification framework developed by Liang et al. (2024), and produces an AI usage score ranging from 0 to 1. The score estimates the proportion of text substantially modified by LLMs, beyond basic spelling and grammatical corrections. The AI usage score should be interpreted as an indirect indicator of AI-assisted writing rather than a direct measure of actual AI use. Higher scores may also reflect differences in authors' English-language proficiency, professional language editing, or discipline-specific writing conventions. These factors may overlap with AI-associated linguistic patterns and introduce measurement uncertainty. We therefore interpret the score cautiously when examining its associations with team characteristics and scholarly visibility.

In Supplementary Figure S1, the distributions of AI usage scores in the PLoS family and the Nature Portfolio are shown. Most publications have relatively low scores, while a smaller proportion have higher scores. Fig. 1 shows the temporal trends in AI usage scores across the two publishing portfolios. Approximately six months after the launch of ChatGPT, June 2023, AI usage scores increased sharply in both portfolios, with a substantially larger increase observed in PLoS publications than in Nature Portfolio publications. Notably, AI usage scores remained low but nonzero before the launch of ChatGPT. This may reflect the use of earlier AI-assisted writing tools or linguistic patterns unrelated to AI assistance that were nevertheless captured by the measurement approach.

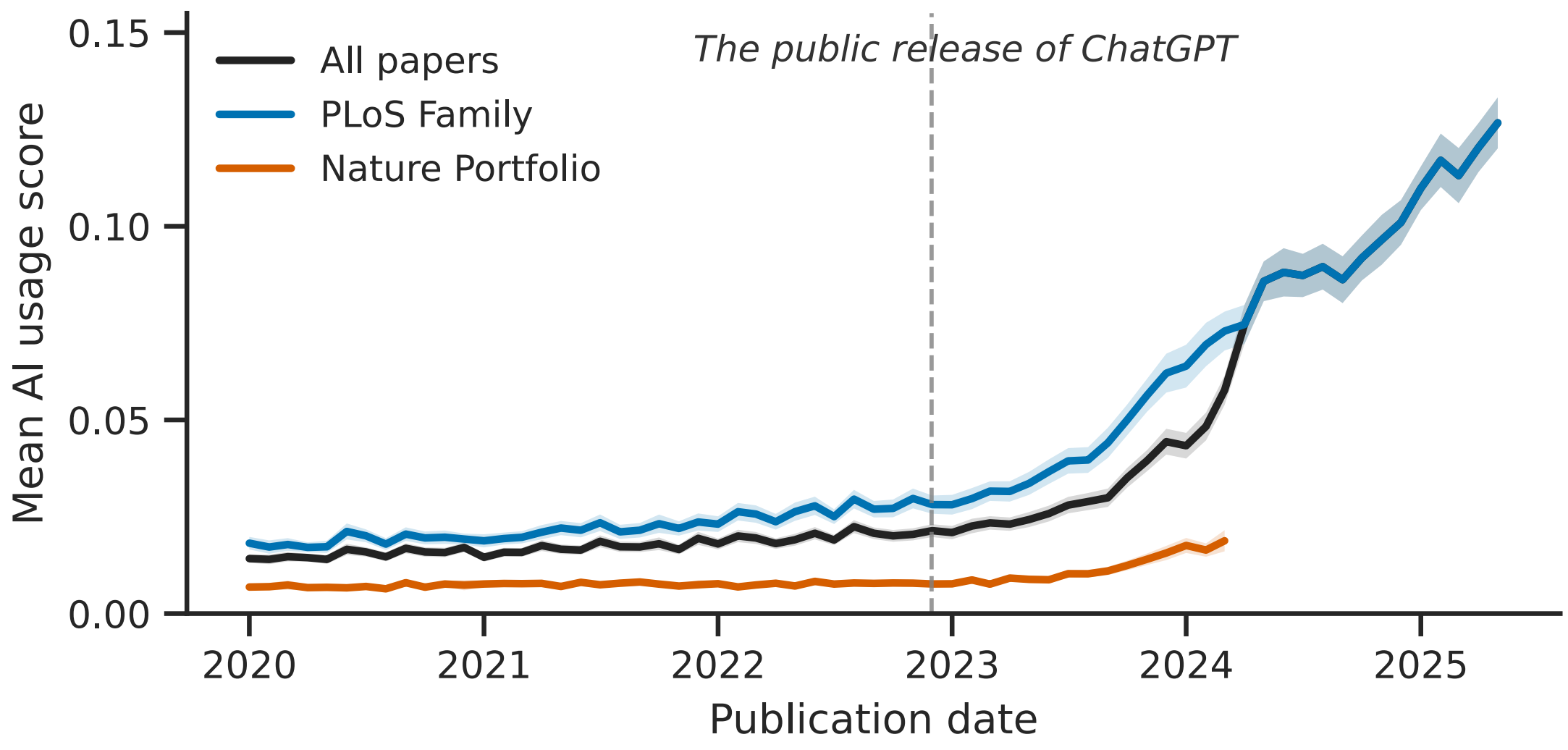

**Figure 1. The temporal trend of AI usage score**

Note: The shade indicates 95% CI.

Based on publication dates and AI usage scores, for descriptive comparisons, we classified publications into three groups: Pre-ChatGPT, Post-ChatGPT (Below Threshold), and Post-ChatGPT (Above Threshold).

**Group 1: Pre-ChatGPT**. This group comprises publications published before November 30, 2022, the date of public availability of ChatGPT.

To distinguish publications with relatively higher AI usage scores from those with lower scores, we used the pre-ChatGPT publication cohort as a reference distribution. Specifically, for each publishing portfolio, we set the classification threshold at the 95th percentile of the AI usage score distribution among publications published before November 30, 2022. This approach assumes that the pre-ChatGPT cohort provides a useful reference for the distribution of scores in the absence of widespread ChatGPT use, although earlier AI writing and language-assistance tools may also have been used. The resulting thresholds were 0.092 for the PLoS family and 0.026 for the Nature Portfolio. We further generate two other groups:

**Group 2: Post-ChatGPT (Below Threshold)**. This group consists of publications published on or after November 30, 2022, with scores below the thresholds.

**Group 3: Post-ChatGPT (Above Threshold)**. This group includes those with scores at or above the thresholds.

The three groups were constructed to compare publications across different periods and levels of estimated AI-modified content. The Post-ChatGPT (Above Threshold) group represents publications with relatively strong AI-modified text signals, whereas the Post-ChatGPT (Below Threshold) group represents publications with lower scores relative to the pre-ChatGPT reference distribution.

The three publication groups were used for descriptive comparisons and propensity score matching, whereas the continuous score was used in the regression analyses. The PLoS family dataset includes 54,999 Pre-ChatGPT publications, 37,069 Post-ChatGPT (Below Threshold) publications, and 11,899 Post-ChatGPT (Above Threshold) publications. The Nature Portfolio dataset includes 26,380, 13,002, and 1,581 publications in the respective groups.

#### *3.2.2 Team academic age*

We measure team academic age using the average career age of the authors listed on a publication (Liu, Jaiswal, et al., 2022; Xu et al., 2022). We used career age as a proxy for authors' academic seniority. For each author, career age was calculated as the number of years between their first publication recorded in OpenAlex and the publication year of the focal paper, inclusive of both years:

$$CareerAge_{a,p} = PublicationYear_i - FirstYear_a + 1 \quad (1)$$

Where $FirstYear_a$ denotes the year when author $a$ published the first publication recorded in OpenAlex, and $PublicationYear_i$ denotes the publication year of focal publication $i$. For example, if an author's first recorded publication was in 2022, this author's career age for a paper published in 2024 would be three years (=2024 − 2022 + 1).

We calculated team academic age $TeamAge_i$ as the arithmetic mean of the career ages of all authors listed on publication $i$. This measure captures the average length of authors' recorded publication careers, rather than their chronological age. Its accuracy therefore depends on the completeness of authors' publication records and the reliability of author disambiguation in OpenAlex.

In addition to examining team academic age as a continuous measure, we classified authors into four career-stage groups, following Robinson-Garcia et al. (2020): (1) junior, with a career age of less than 5 years; (2) early-career, with a career age of 5 to less than 15 years; (3) mid-career, with a career age of 15 to less than 30 years; and (4) late-career, with a career age of 30 years or more. These categories were used to describe the career-stage composition of publication teams.

#### *3.2.3 Publication team size*

Following established practice (Lee et al., 2015; Liu, Jaiswal, et al., 2022), we measured publication team size as the total number of authors listed in a publication's byline. This measure captures the size of the author team associated with a publication and does not necessarily reflect the number of individuals involved in the broader research project.

*3.2.4 Publication scientific visibility*

We measured scientific visibility as the number of citations received by a publication within one year of its publication date. The OpenAlex dataset was updated in July 2026, and all focal publications in our sample were published by May 2025, allowing for a complete one-year citation window. As a robustness check, we also measured citations accumulated within 180 days of publication, approximately half a year. The corresponding results are reported in Supplementary Table S7, S9 and Figure S8.

## 3.3 Methods

*3.3.1 Statistical analyses*

To address RQ1–RQ3, we applied statistical analyses to compare team academic age, publication team size and one-year citations across the three publication groups. We used the Welch's t-tests to assess pairwise differences in the mean of team academic age and publication team size, which were non-normally distributed according to the Kolmogorov–Smirnov tests (team academic age: $D = 0.141$, $p < 0.01$; publication team size: $D = 0.031$, $p < 0.01$). To investigate publication team size, we also consider the career age composition, and author positions. To examine differences in short-term scientific visibility (RQ3), we compared citation counts accumulated within one year of publication across the three groups.

*3.3.2 Multivariate regression analyses*

To account for observable factors that may be associated with both AI usage scores and team characteristics and short-term scientific visibility, we estimated multivariate regression models, including ordinary least squares (OLS), quantile regression, and Poisson regression.

We restricted the main regression analyses to publications published after November 30, 2022, including both the Post-ChatGPT (Below Threshold) and Post-ChatGPT (Above Threshold) groups. We estimated the models using the pooled sample and separately for the PLoS family and the Nature Portfolio.

In the multivariate regression analyses, we used the continuous AI usage score as the main explanatory variable. This approach preserves variation in estimated AI-modified content across publications and allows us to examine how differences in the score are associated with publication team characteristics, and short-term citations, conditional on observed covariates.

To address RQ1, we estimated team academic age using Equation 2.

$$TeamAge_i = \beta_0 + \beta_1 AIUsageScore_i + \Gamma \boldsymbol{X_i} + Time_i + Discipline_i + Journal_i + \varepsilon_i \quad (2)$$

Let $i$ index publications. The dependent variable, $TeamAge_i$, is the average career age of the authors listed on publication $i$. The main explanatory variable, $AI\ Usage\ Score_i$, measures the estimated extent of AI-modified content in the publication's full text. The vector $\boldsymbol{X_i}$ contains observed author-level and publication-level control variables that may be associated with publication team characteristics.

To account for the prior experience and collaboration patterns of leading authors, we included six controls for the first author and six for the last author. For each author, these controls capture (1) the number of publications in the preceding three years, (2) the average cumulative citations per publication over the preceding three years, (3) the number of unique collaborators over the preceding three years, (4) the average team size of their publications over the preceding three years, (5) the average team academic age of their publications over the preceding three years, and (6) English-speaking country, a binary variable of whether the author's affiliated country has an English-speaking population of at least 20%. These measures capture aspects of authors' prior productivity, citations, collaborative networks, publication team characteristics, and language environment (Liu et al., 2025; Nasra & Oliver, 2025). The classification of English-speaking countries was based on the CEPII GeoDist database[6] (Mayer & Zignago, 2011), with the list of countries provided in Supplementary Table S11. All continuous control variables were log-transformed using ln(x+1) to reduce skewness. $\varepsilon_i$ denotes the error term.

We further included fixed effects for publication year-month, discipline, and journal to account for temporal trends and systematic differences across disciplines and journals. The model was estimated using OLS, with standard errors clustered by journal and publication year-month to account for potential dependence among observations within these groups.

We employed quantile regression models to explore if the relationship between team academic age and AI usage score is not uniform across the seniority spectrum. This approach offers a more comprehensive statistical analysis opportunity compared to the traditional mean regression model like OLS (Koenker & Hallock, 2001; Liu et al., 2024). Quantile regression allows estimation of the relationship between explanatory variables and the conditional quantity of the dependent variable without assuming a specific conditional distribution.

We conducted quantile regression analyses at the 25th, 50th, and 75th percentiles. These percentiles capture differences among publications with relatively younger, middle-aged, and older author teams, respectively. This approach allows us to assess whether the association between AI usage score and team academic age differs across the distribution, rather than focusing solely on the conditional mean. The quantile regression model is specified in Equation 3.

[6] https://www.cepii.fr/cepii/en/bdd_modele/bdd_modele_item.asp?id=6

$$Q_{yi}(\tau) = \sum_{m=1}^{k} \beta_{\tau,m} x_{im} \quad (3)$$

Where $\beta_{\tau,m}$ is an unknown parameter, and $Q_{yi}(\tau)$ denotes the $\tau^{th}$ conditional quantile of $TeamAge_i$ that is denoted by $yi$ (0< $\tau$ <1), i.e., the $\tau^{th}$ quantile of $yi$ given $x_i$. Unlike the conditional mean function estimated by OLS, the quantile regression coefficients are estimated by minimizing an asymmetric absolute loss function, which weights positive and negative residuals differently depending on the chosen quantile $\tau$.

To address RQ2, we estimated publication team size, a discrete count variable, using Poisson regression. Publication team size of $i$ is estimated by Equation 4.

$$TeamSize_i = \beta_0 + \beta_1 AIUsageScore_i + \Gamma \boldsymbol{X_i} + Time_i + Discipline_i + Journal_i + \varepsilon_i (4)$$

Where $TeamSize_i$ refers to the key dependent variable, measured by the number of authors of publication $i$; all other specifications follow those shown in Equation 2. $\beta_1$ reflects the estimated relationship between a publication's AI usage score and its team size. To facilitate a more intuitive interpretation, we report the incidence rate ratio (IRR), calculated as the exponentiation of $\beta_1$, which indicates the multiplicative change in publication team size associated with a one-unit increase in the AI usage score.

To address RQ3, we applied OLS regression models to estimate whether AI usage score is associated with the one-year citation count (log-transformed).

$$ln(Citation_i + 1) = \beta_0 + \beta_1 AIUsageScore_i + \Gamma \boldsymbol{X_i} + Time_i + Discipline_i + Journal_i + \varepsilon_i (5)$$

In this model, publication team size and team academic age are considered as controls variables; all other specifications follow those presented in Equation 2.

To address RQ4, we introduced interaction terms between AI usage score and team size, and between AI usage score and team academic age, into the regression models predicting short-term citation counts. These interaction terms allow us to examine whether the associations of team size and academic age with citation visibility vary across different levels of AI usage. The summary statistics of variables employed in this study are shown in Supplementary Table S1.

### *3.3.3 Propensity score matching*

To assess the robustness of our findings to observable differences between publications, we conducted propensity score matching (PSM). We matched Post-ChatGPT (Above Threshold) with Post-ChatGPT (Below Threshold) within the same journal and discipline, using author-level characteristics measured from prior publication records. Matching improves comparability on observed covariates (Caliendo & Kopeinig, 2008; Liu & Hu, 2022).

We conducted PSM separately for each dependent variable to compare publications with higher and lower AI usage scores. The treatment group consisted of publications in the Post-ChatGPT (Above Threshold) group, while the control group consisted of publications in the Post-ChatGPT (Below Threshold) group. To improve comparability, matching was conducted within the same journal and discipline. We first estimated each publication's propensity score, defined as the conditional probability of belonging to the Above-Threshold group, using logistic regression. The covariates followed the control variables specified for each dependent variable in Equations 2, 4, and 5. Specifically, for team size and team academic age, we included twelve author-level controls. For citation counts, we additionally controlled for publication-team size and team academic age. Before matching, we imposed a common support condition (min-max overlap), restricting the sample to observations whose propensity scores fell within the overlapping range of the two publication groups. We then applied 1:1 nearest-neighbor matching without replacement, pairing each Above-Threshold publication with a Below-Threshold publication with the closest propensity score. We assessed post-matching covariate balance using Standardized Mean Difference (SMD) between the Above-Threshold and Below-Threshold groups. As reported in Supplementary Tables S3 and S4 in the Appendix, matching substantially reduced covariate differences between the two groups. SMDs of all covariates are smaller than 0.1. Finally, we calculated the matched mean difference (MMD) for each outcome as the difference in mean values between the matched treatment and control groups.

The matching procedure was intended to improve comparability between the two groups with respect to observed covariates. Nevertheless, PSM cannot account for unobserved differences between publications. Accordingly, the matched mean differences are interpreted as conditional associations rather than causal effects.

# 4 Results

## 4.1 The relationship between AI usage score and team academic age

### *4.1.1 Mean and distribution of team academic age*

Publications in the Post-ChatGPT (Above Threshold) group exhibit lower mean team academic age than those in the other two groups (Fig. 2). In the PLoS family, the mean team academic age is 17.26 years for the Above-Threshold group, compared with 19.58 years for the Below-Threshold group and 19.87 years for the Pre-ChatGPT group. The corresponding means in the Nature Portfolio are 20.59, 21.36, and 21.21 years, respectively. Pairwise Welch's t-tests further indicate statistically significant differences in the both portfolio groups.

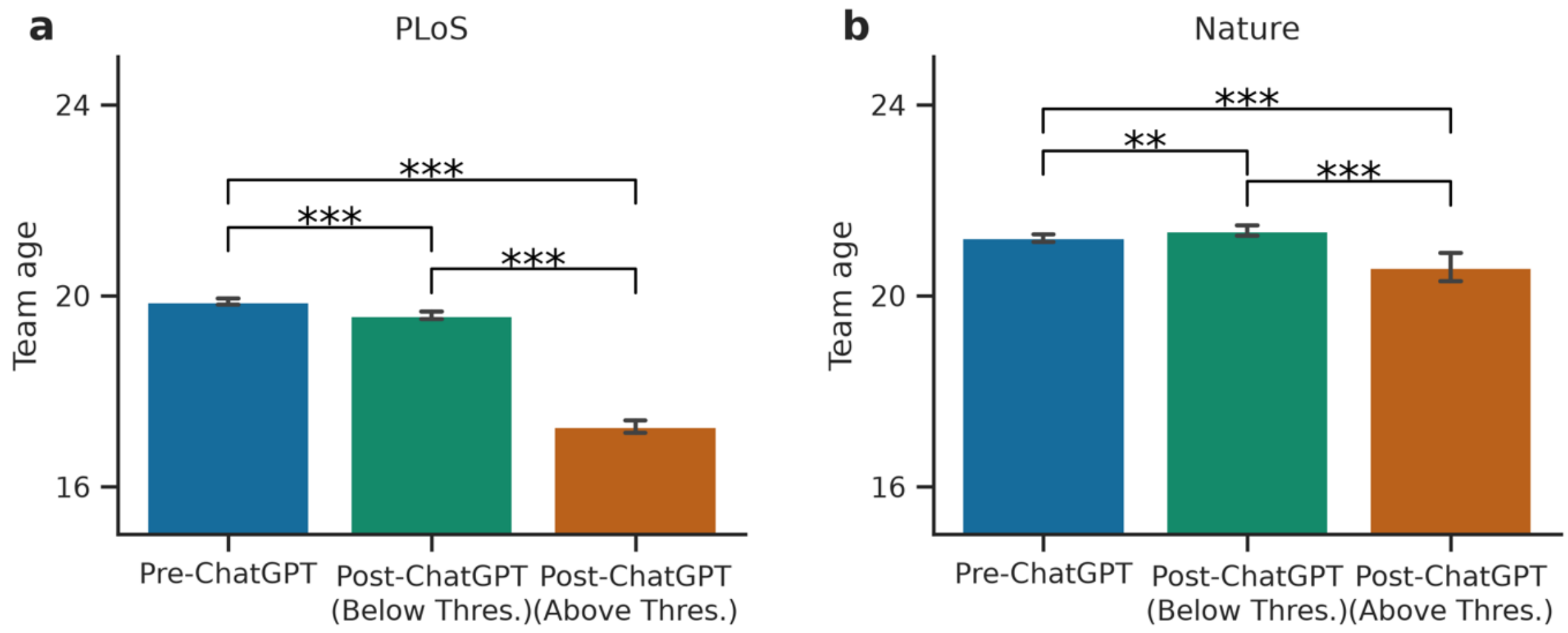

**Figure 2. Team academic age across three publication groups.**

Note: Panels (a) and (b) compare mean team academic age across three publication groups for the PLoS family and Nature Portfolio, respectively. Pairwise group differences are assessed using Welch's t-tests. * $p < 0.1$, ** $p < 0.05$, *** $p < 0.01$.

#### *4.1.2 Academic team age composition*

We examined differences in team career-age composition across the three groups. In the PLoS family, the Above-Threshold group has a higher representation of researchers with fewer than 10 years of career experience and a lower representation of those with more than 25 years of experience (Fig. 3a). This pattern is also reflected in career-stage categories, with more junior and early-career researchers and fewer late-career researchers in this group (Fig. 3b–c).

The pattern differs in the Nature Portfolio. The Above-Threshold group has a higher representation of researchers with 20–30 years of career experience and a lower representation of those with more than 40 years of experience (Fig. 3d). Accordingly, this group includes more mid-career and fewer late-career researchers, while the proportions of junior and early-career researchers remain broadly similar across the three groups (Fig. 3e–f). Overall, the differences in team career-age composition are more broadly distributed across career stages in the PLoS family, but are concentrated among mid- and late-career researchers in the Nature Portfolio.

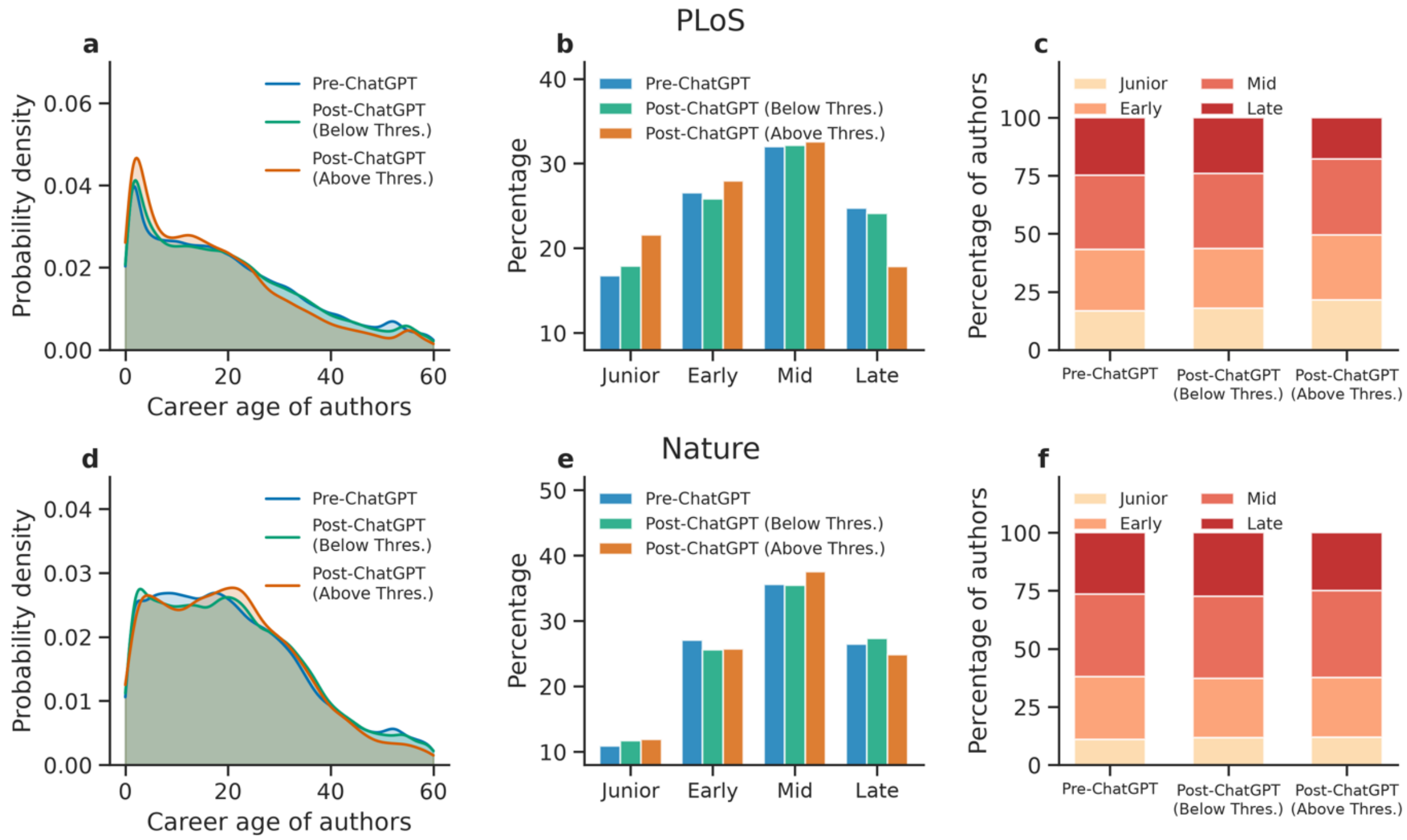

**Figure 3. Career age composition of publication teams across the three publication groups.**

Note: (a, d): The Kernel density estimation (KDE) graph of authors' career age. (b, e): Percentage distribution of authors across four career stages. (c, f): The proportion of authors in the four career stages across three groups.

### *4.1.3 Authorship positions*

We examined differences in authors' career age across first, last, and middle authorship positions. In the PLoS family, the Above-Threshold group has a lower mean career age across all three positions than the Pre-ChatGPT and Below-Threshold groups (Fig. 4a). For example, the mean career age of first authors is 14.2 years, compared with 15.4 and 14.9 years in the two comparison groups, respectively. A similar but less pronounced pattern is observed in the Nature Portfolio (Fig. 4b). Overall, the lower career age associated with the Above-Threshold group extends across authorship positions in both portfolios.

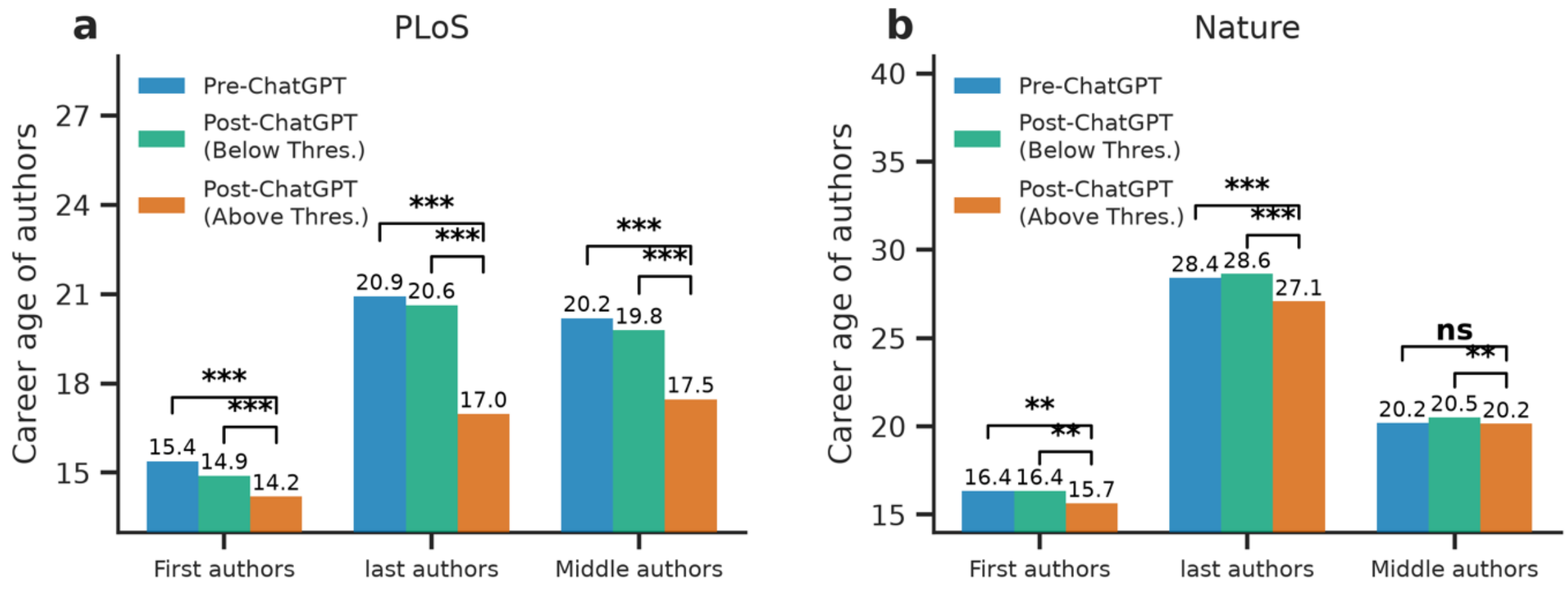

**Figure 4. Authors' career age and authorship positions across the three publication groups.**

Note: Career age is compared across the three publication groups for first, last, and middle authors within PLoS family (a) and Nature portfolio (b) respectively. Pairwise differences are assessed using the Welch's t-tests. * $p < 0.1$, ** $p < 0.05$, *** $p < 0.01$.

### *4.1.4 Regression analyses and propensity score matching*

The OLS regression results show a significant negative association between AI usage score and team academic age in both publishing portfolios (Table 1). For the PLoS family, the coefficient on AI usage score is $-2.566$ ($p < 0.01$), indicating that publications with higher AI usage scores tend to have teams with lower average author career ages. Specifically, a 0.01 increase in AI usage score is associated with a 0.026-year decrease in average team career age. The association is stronger in the Nature Portfolio, where the coefficient is $-6.855$ ($p < 0.01$). A 0.01 increase in AI usage score is associated with a 0.069-year decrease in average team career age.

**Table 1. OLS regression results for the estimated relationship between AI usage score and team academic age based on the publications published after Nov. 2022.**

| Model | (1) | (2) | (3) |
|---|---|---|---|
| Group | Overall | PLoS | Nature |
| AI usage score | -2.487*** | -2.566*** | -6.855** |
| | (0.270) | (0.230) | (2.672) |
| First Author Publication Count (ln) | 0.313*** | 0.403*** | -0.065 |
| | (0.077) | (0.076) | (0.082) |
| First Author Collaborator Count (ln) | 0.047 | -0.084 | 0.545*** |
| | (0.125) | (0.067) | (0.067) |
| First Author Prior Team Size (ln) | -0.006 | 0.019 | -0.053 |
| | (0.037) | (0.032) | (0.090) |
| First Author Prior Team Age (ln) | 8.415*** | 8.173*** | 10.613*** |
| | (0.265) | (0.091) | (0.403) |
| First Author Previous Avg. Impact (ln) | 0.020 | 0.037* | 0.003 |
| | (0.024) | (0.018) | (0.037) |
| First Author English Speaking = True | -0.208 | -0.053 | -0.696*** |
| | (0.185) | (0.106) | (0.127) |
| Last Author Publication Count (ln) | 0.607*** | 0.653*** | 0.424*** |
| | (0.122) | (0.109) | (0.093) |
| Last Author Collaborator Count (ln) | -0.377*** | -0.420*** | -0.257* |
| | (0.099) | (0.087) | (0.125) |
| Last Author Prior Team Size (ln) | 0.128 | 0.124 | 0.001 |
| | (0.093) | (0.108) | (0.202) |
| Last Author Prior Team Age (ln) | 5.239*** | 5.220*** | 6.290*** |
| | (0.042) | (0.059) | (0.342) |
| Last Author Previous Avg. Impact (ln) | -0.293*** | -0.264*** | -0.260*** |
| | (0.040) | (0.061) | (0.086) |
| Last Author English Speaking = True | 0.030 | -0.023 | 0.230** |
| | (0.112) | (0.106) | (0.095) |
| Fixed Effects | | | |
| Year-Month | Yes (30 period) | Yes (30 periods) | Yes (16 periods) |
| Discipline | Yes (19 fields) | Yes (19 fields) | Yes (17 fields) |
| Journal | Yes (29 venues) | Yes (14 venues) | Yes (15 venues) |
| Model Fit | | | |
| Adjusted $R^2$ | 0.432 | 0.443 | 0.326 |
| Obs. | 63,551 | 48,968 | 14,583 |

Note: Coefficients are unstandardized estimates. Cluster standard errors (Journal + Year-Month) are reported in parentheses. Significance levels: * $p < 0.1$, ** $p < 0.05$, *** $p < 0.01$. The AI usage score is a continuous variable ranging from 0 to 1.

In the matched PLoS sample, the mean team academic age is 17.21 years in the Above-Threshold group, significantly lower than the 17.81 years observed in the Below-Threshold group (Table 2). A similar difference is found in the Nature Portfolio, where the corresponding means are 20.91 and 21.52 years, respectively. The MMD are negative and statistically significant in both portfolios, indicating that the differences in team academic age persist after matching on the observed covariates.

We further examined whether the association between AI usage score and team academic age varies across the team-age distribution using quantile regression at the 25th, 50th, and 75th percentiles (Supplementary Table S2). The results indicate that the association is not uniform across the distribution. In the PLoS family, the negative coefficient is largest in magnitude at the 25th percentile ($-3.059$, $p < 0.01$), suggesting a stronger association among teams with lower academic age. The Nature Portfolio shows a similar pattern, with the largest negative coefficient at the 25th percentile ($-7.116$, $p < 0.01$). These findings suggest that the negative association between AI usage score and team academic age is more pronounced among younger teams in both publishing portfolios.

**Table 2. PSM analyses of team academic age, publication team size, and one-year citation counts between the Post-ChatGPT (Above Threshold) and Post-ChatGPT (Below Threshold) groups after Nov. 2022.**

| **Dependent Var.** | Retained Pairs by Common Support | Post-ChatGPT (Above Threshold) Mean | Post-ChatGPT (Below Threshold) Mean | Matched Mean Difference (MMD) | p-value |
|---|---|---|---|---|---|
| **Panel A: Overall (13,480 pairs in total)** | | | | | |
| Team Age | 13,475 | 17.644 | 18.495 | -0.851*** | < 0.01 |
| Team Size | 13,475 | 6.127 | 6.388 | -0.261*** | < 0.01 |
| One-year citation (ln) | 13,469 | 1.124 | 0.970 | 0.154*** | < 0.01 |
| **Panel B: PLoS (11,899 pairs in total)** | | | | | |
| Team Age | 11,892 | 17.211 | 17.810 | -0.599*** | < 0.01 |
| Team Size | 11,892 | 5.811 | 6.065 | -0.254*** | < 0.01 |
| One-year citation (ln) | 11,888 | 0.983 | 0.817 | 0.166*** | <0.01 |
| **Panel C: Nature (1,581 pairs in total)** | | | | | |
| Team Age | 1,581 | 20.912 | 21.521 | -0.608*** | <0.01 |
| Team Size | 1,581 | 8.501 | 8.756 | -0.256 | >0.01 (0.108) |
| One-year citation (ln) | 1,581 | 2.186 | 2.092 | 0.093*** | <0.01 |

Note: MMD denotes the mean difference between Above-Threshold and Below-Threshold publications within matched pairs. * p < 0.1, ** p < 0.05, *** p < 0.01.

### 4.2 The relationship between AI usage score and team size

Our analyses indicate that publications in the Post-ChatGPT (Above Threshold) group tend to have smaller teams than those in the two comparison groups. This pattern is observed in both publishing portfolios, although the strength of the statistical evidence varies across analyses.

The descriptive comparisons show that the Post-ChatGPT (Above Threshold) group has a lower mean team size than the other two groups in both datasets (Fig. 5a and 5b). The Poisson regression results further support this association. For the PLoS family, the coefficient on AI usage score is $-0.239$ ($p < 0.01$), corresponding to an IRR of 0.787. This estimate indicates that a one-unit increase in AI usage score, from 0 to 1, is associated with a 21.3% lower expected team size, holding the other variables constant. The association is stronger in the Nature Portfolio, where the estimated coefficient is $-0.484$ ($p < 0.1$), corresponding to a 0.616 IRR and a 38.4% lower expected team size for a one-unit increase in AI usage score.

The PSM results are broadly consistent with these findings (Table 2). In the matched PLoS sample, the mean team size is significantly smaller in the Post-ChatGPT (Above Threshold) group than in the Post-ChatGPT (Below Threshold) group (5.81 versus 6.07). In the matched Nature sample, the corresponding means are 8.50 and 8.75 for the Post-ChatGPT (Above Threshold) group than in the Post-ChatGPT (Below Threshold) group, respectively. The mean team size is slightly smaller in the Post-ChatGPT (Above Threshold) group, while it is insignificant ($p = 0.108$).

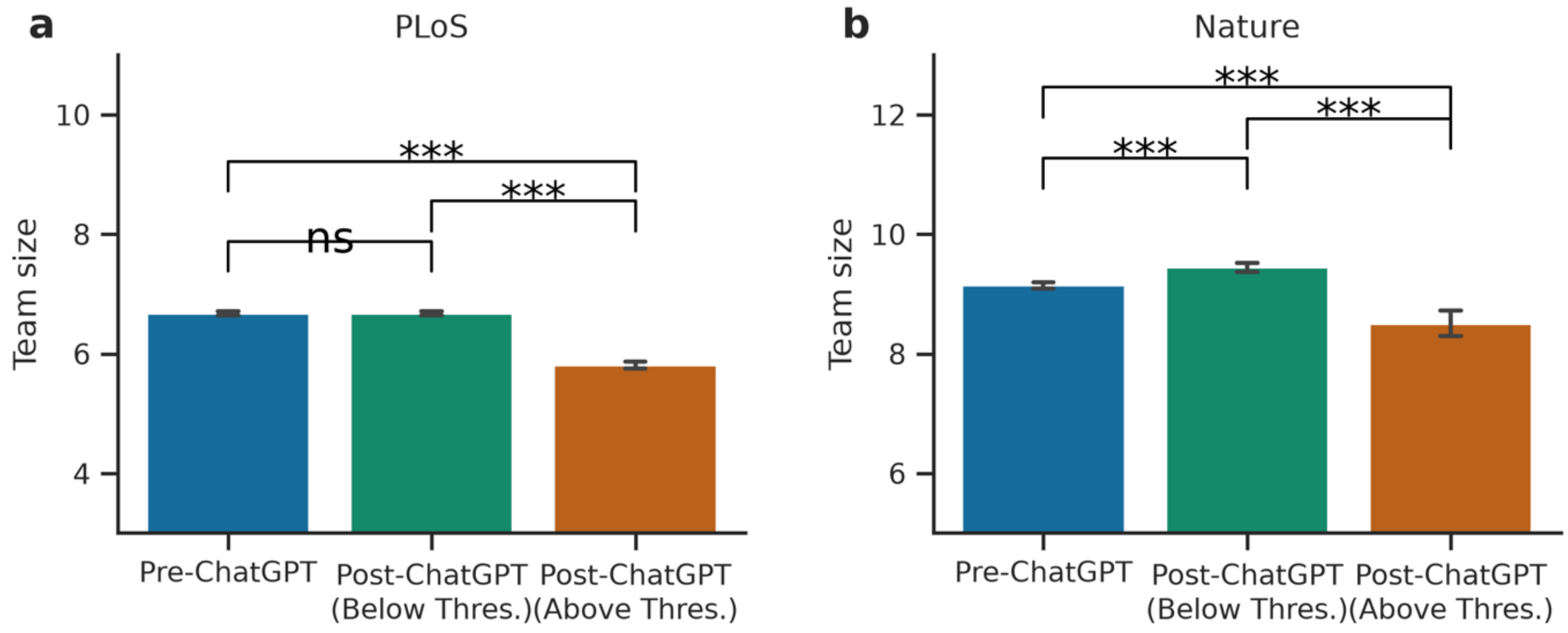

**Figure 5. AI usage scores and team size across three publication groups.**

Note: Bar plots for the PLoS family (a) and Nature portfolio (b) demonstrate that the Post-ChatGPT (Above Threshold) group (orange bars) has a significantly smaller average team size compared to the other two groups (Welch's t-tests: *p < 0.1, **p < 0.05, ***p < 0.01, with ns indicating not significant).

**Table 3. Poisson regression results for the estimated relationship between AI usage score and publication team size after Nov. 2022.**

| Model | (1) | | (2) | | (3) | |
|---|---|---|---|---|---|---|
| Group | Overall | | PLoS | | Nature | |
| | Coeff. | IRR | Coeff. | IRR | Coeff. | IRR |
| AI usage score | -0.252*** | 0.777 | -0.239*** | 0.787 | -0.484* | 0.616 |
| | (0.044) | | (0.042) | | (0.290) | |
| First Author Publication Count (ln) | -0.229*** | 0.795 | -0.229*** | 0.796 | -0.230*** | 0.794 |
| | (0.008) | | (0.010) | | (0.005) | |
| First Author Collaborator Count (ln) | 0.277*** | 1.319 | 0.276*** | 1.318 | 0.281*** | 1.324 |
| | (0.007) | | (0.009) | | (0.006) | |
| First Author Prior Team Size (ln) | 0.040*** | 1.041 | 0.045*** | 1.046 | 0.028*** | 1.028 |
| | (0.008) | | (0.007) | | (0.006) | |
| First Author Co-author Avg. Career Age (ln) | -0.010*** | 0.905 | -0.091*** | 0.913 | -0.169*** | 0.844 |
| | (0.011) | | (0.002) | | (0.013) | |
| First Author Previous Avg. Impact (ln) | -0.035*** | 0.966 | -0.040*** | 0.961 | -0.025*** | 0.976 |
| | (0.004) | | (0.002) | | (0.003) | |
| First Author English Speaking = True | -0.006 | 0.994 | 0.000 | 1.000 | -0.022* | 0.978 |
| | (0.009) | | (0.008) | | (0.013) | |
| Last Author Publication Count (ln) | -0.124*** | 0.884 | -0.132*** | 0.877 | -0.102*** | 0.903 |
| | (0.007) | | (0.004) | | (0.006) | |
| Last Author Collaborator Count (ln) | 0.186*** | 1.204 | 0.198*** | 1.218 | 0.149*** | 1.161 |
| | (0.011) | | (0.004) | | (0.007) | |
| Last Author Prior Team Size (ln) | 0.049*** | 1.050 | 0.046*** | 1.047 | 0.062*** | 1.064 |
| | (0.003) | | (0.003) | | (0.003) | |
| Last Author Prior Team Age (ln) | -0.024*** | 0.976 | -0.031*** | 0.970 | -0.027 | 0.974 |
| | (0.005) | | (0.001) | | (0.025) | |
| Last Author Previous Avg. Impact (ln) | -0.052*** | 0.949 | -0.052*** | 0.949 | -0.059*** | 0.943 |
| | (0.002) | | (0.003) | | (0.004) | |
| Last Author English Speaking = True | -0.011 | 0.989 | -0.007 | 0.993 | -0.018 | 0.982 |
| | (0.007) | | (0.005) | | (0.012) | |
| Fixed Effects | | | | | | |
| Year-Month | Yes (30 periods) | | Yes (30 periods) | | Yes (16 periods) | |
| Discipline | Yes (19 fields) | | Yes (19 fields) | | Yes (17 fields) | |
| Journal | Yes (29 venues) | | Yes (14 venues) | | Yes (15 venues) | |
| Model Fits | | | | | | |
| Pseudo $R^2$ | 0.161 | | 0.135 | | 0.127 | |
| Obs. | 63,551 | | 48,968 | | 14,583 | |

Note: Coefficients are unstandardized Poisson log-counts. Standard errors are clustered by journal and year-month, and are reported in parentheses. Significance levels: * p < 0.1, ** p < 0.05, *** p < 0.01. The AI usage score is a continuous variable ranging from 0 to 1.

### 4.3 The relationship between AI usage score and short-term citations

For the PLoS family, the mean number of one-year citations is approximately 3.54 for the Pre-ChatGPT group, 2.28 for the Post-ChatGPT (Below Threshold) group, and 2.72 for the Post-ChatGPT (Above Threshold) group (Fig. 6a). Pairwise comparisons indicate that the Above-Threshold group therefore receives more citations than the Below-Threshold group.

A similar pattern emerges in the Nature Portfolio (Fig. 6b). The mean citation count is approximately 14.48 for the Pre-ChatGPT group, 12.09 for the Post-ChatGPT (Below Threshold) group, and 12.74 for the Post-ChatGPT (Above Threshold) group. Although the mean citation count for the Above-Threshold group is slightly higher than that of the Below-Threshold group, this unconditional difference is not statistically significant (p=0.254).

Notably, after controlling for author characteristics, publication timing, discipline, and journal fixed effects, the results show a positive and statistically significant association between AI usage score and one-year citations in the overall sample and in both publishing portfolios (Table 4). For the full sample, the estimated coefficient is 0.760 ($p < 0.01$). The corresponding coefficients are 0.779 ($p < 0.01$) for the PLoS family and 1.529 ($p < 0.01$) for the Nature Portfolio. These estimates indicate that a 0.1 increase in the score is associated with 7.9%, 8.1%, and 16.5% additional citations, respectively, conditional on the included covariates and fixed effects. Overall, the regression results suggest that the positive association between AI usage score and short-term citation is robust and highly significant.

In the overall matched sample, the mean log-transformed citation count is 1.124 for the Above-Threshold group, compared with 0.970 for the Below-Threshold group (Table 2). The MMD is 0.154 ($p < 0.01$). The difference is also statistically significant in the PLoS family, where the corresponding means are 0.983 and 0.817 citations, yielding an MMD of 0.166 ($p < 0.01$). For the Nature Portfolio, the mean citation count is 2.186 in the above-threshold group and 2.092 in the below-threshold group. The MMD is positive (0.093) and significant ($p < 0.01$).

The adjusted regression and matched-sample results consistently indicate a positive association between AI usage score and short-term citations, although the unconditional descriptive difference is not statistically significant in the Nature Portfolio.

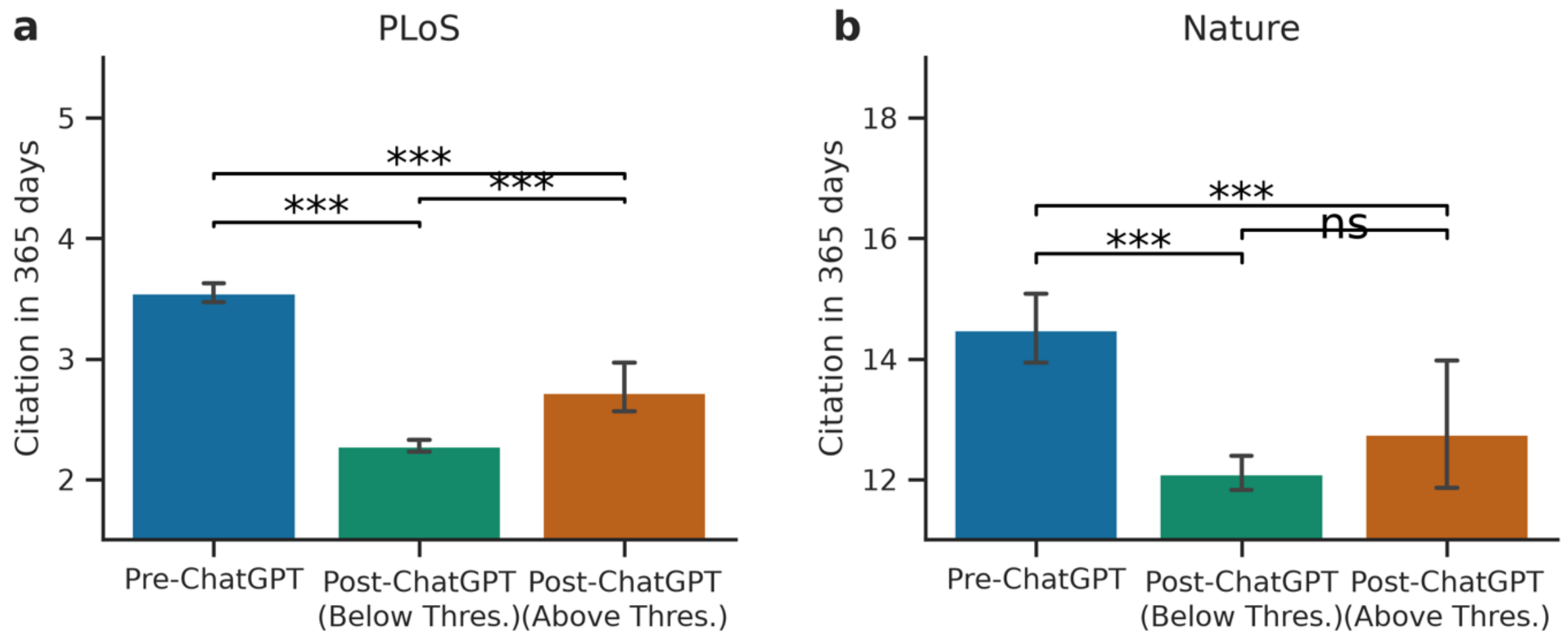

**Figure 6. The average citation counts in one year across three groups.**

Note: Bar plots for the PLoS family (a) and Nature portfolio (b) illustrate the differences in citation counts in one-year time window across the three groups. The Welch's t-tests results (*p < 0.1, p < 0.05, ***p < 0.01) show that: The Post-ChatGPT (Above Threshold) group shows a higher mean citation count than the Below Threshold group in both portfolios, although this difference is statistically significant only for the PLoS family.

**Table 4. OLS regression results for the estimated relationship between AI usage score and the one-year citations after Nov. 2022.**

| Model | (1) | (2) | (3) |
|---|---|---|---|
| Group | Overall | PLoS | Nature |
| AI usage score | 0.760*** | 0.779*** | 1.529*** |
| | (0.046) | (0.036) | (0.196) |
| Team Age | -0.003*** | -0.003*** | -0.005*** |
| | (0.000) | (0.000) | (0.001) |
| Team Size | 0.017*** | 0.013*** | 0.023*** |
| | (0.003) | (0.001) | (0.002) |
| First Author Publication Count (ln) | 0.041*** | 0.031*** | 0.075*** |
| | (0.013) | (0.007) | (0.005) |
| First Author Collaborator Count (ln) | -0.000 | 0.003 | -0.013* |
| | (0.005) | (0.004) | (0.007) |
| First Author Prior Team Size (ln) | -0.096*** | -0.073*** | -0.136*** |
| | (0.019) | (0.004) | (0.021) |
| First Author Prior Team Age (ln) | -0.014* | -0.010 | -0.048 |
| | (0.008) | (0.007) | (0.045) |
| First Author Previous Avg. Impact (ln) | 0.080*** | 0.059*** | 0.108*** |
| | (0.016) | (0.004) | (0.012) |
| First Author English Speaking = True | -0.017 | 0.010** | -0.100*** |
| | (0.021) | (0.004) | (0.015) |
| Last Author Publication Count (ln) | 0.034*** | 0.026*** | 0.050*** |
| | (0.010) | (0.008) | (0.008) |
| Last Author Collaborator Count (ln) | -0.003 | -0.005 | 0.039** |
| | (0.009) | (0.004) | (0.014) |
| Last Author Prior Team Size (ln) | -0.095*** | -0.065*** | -0.184*** |
| | (0.027) | (0.004) | (0.028) |
| Last Author Prior Team Age (ln) | -0.053*** | -0.046*** | -0.027 |
| | (0.009) | (0.007) | (0.018) |
| Last Author Previous Avg. Impact (ln) | 0.104*** | 0.075*** | 0.220*** |
| | (0.028) | (0.004) | (0.021) |
| Last Author English Speaking = True | 0.017*** | 0.010** | 0.055** |
| | (0.006) | (0.004) | (0.025) |
| Fixed Effects | | | |
| Year-Month | Yes (30 periods) | Yes (30 periods) | Yes (16 periods) |
| Discipline | Yes (19 fields) | Yes (19 fields) | Yes (17 fields) |
| Journal | Yes (29 venues) | Yes (14 venues) | Yes (15 venues) |
| Model Fits | | | |
| Adj. $R^2$ | 0.426 | 0.094 | 0.369 |
| Obs. | 63,551 | 48,968 | 14,583 |

Note: Coefficients are unstandardized estimates. Standard errors are clustered by journal and year-month, and are reported in parentheses. Significance levels: * $p < 0.1$, ** $p < 0.05$, *** $p < 0.01$. The AI usage score is a continuous variable ranging from 0 to 1.

## 4.4 The moderation of AI usage score on the relationship between team academic age, publication team size and short-term citation

The results show no statistically significant interaction between AI usage score and team academic age across the overall sample or either publication portfolio (Table 5). This suggests that the association between team academic age and citation visibility does not vary significantly with the level of AI usage. In contrast, the interaction between AI usage score and team size is negative and statistically significant in the overall sample ($\beta = -0.034,\ p < 0.01$) and the Nature portfolio ($\beta = -0.101,\ p < 0.05$), while it is negative and marginally significant in the PLoS portfolio ($\beta = -0.028,\ p < 0.1$). These results indicate that the positive association between team size and citation visibility becomes weaker as AI usage score increases. The pattern is consistent across the two publication portfolios, although the statistical evidence is stronger in the overall sample and Nature portfolio. The estimated interaction of AI usage score on the relationship between team characteristics and short-term citations is shown in Figure 7. Taken together, the findings suggest that AI-assisted writing is associated with a weaker citation visibility advantage for larger research teams, whereas no comparable moderating relationship is observed for team academic age.

**Table 5. OLS regression results for the estimated interaction of AI usage score on the relationship between team academic age, publication team size and the one-year citations after Nov. 2022.**

| Model | (1) | (2) | (3) | (4) | (5) | (6) |
|---|---|---|---|---|---|---|
| Group | Overall | PLoS | Nature | Overall | PLoS | Nature |
| AI usage score | 0.739*** | 0.785*** | 1.205 | 0.932 *** | 0.926*** | 2.368*** |
| | (0.046) | (0.036) | (1.295) | (0.085) | (0.093) | (0.440) |
| Team Age | -0.003*** | -0.003*** | -0.005*** | -0.003*** | -0.003*** | -0.004*** |
| | (0.000) | (0.000) | (0.001) | (0.000) | (0.000) | (0.001) |
| AI usage score * Team Age | **0.001** | **-0.0004** | **0.016** | | | |
| | **(0.003)** | **(0.003)** | **(0.061)** | | | |
| Team Size | 0.017*** | 0.013*** | 0.023*** | 0.018*** | 0.015*** | 0.025*** |
| | (0.003) | (0.001) | (0.002) | (0.003) | (0.001) | (0.002) |
| AI usage score * Team Size | | | | **-0.034***** | **-0.028*** | **-0.101*** |
| | | | | **(0.012)** | **(0.015)** | **(0.048)** |
| First Author Publication Count (ln) | 0.041*** | 0.031*** | 0.075*** | 0.041*** | 0.031*** | 0.075*** |
| | (0.013) | (0.007) | (0.005) | (0.013) | (0.007) | (0.005) |
| First Author Collaborator Count (ln) | -0.000 | 0.003 | -0.013* | -0.000 | 0.003 | -0.013* |
| | (0.005) | (0.004) | (0.007) | (0.005) | (0.004) | (0.007) |
| First Author Prior Team Size (ln) | -0.096*** | -0.073*** | -0.136*** | -0.097*** | -0.074*** | -0.136*** |
| | (0.019) | (0.004) | (0.021) | (0.019) | (0.004) | (0.021) |
| First Author Prior Team Age (ln) | -0.014* | -0.010 | -0.048 | -0.014* | -0.010 | -0.049 |
| | (0.008) | (0.007) | (0.045) | (0.008) | (0.007) | (0.045) |
| First Author Previous Avg. Impact (ln) | 0.080*** | 0.059*** | 0.108*** | 0.080*** | 0.059*** | 0.108*** |
| | (0.016) | (0.004) | (0.012) | (0.016) | (0.004) | (0.013) |
| First Author English Speaking = True | -0.017 | 0.010** | -0.100*** | -0.016 | 0.011** | -0.099*** |
| | (0.021) | (0.004) | (0.015) | (0.021) | (0.004) | (0.015) |
| Last Author Publication Count (ln) | 0.034*** | 0.026*** | 0.050*** | 0.034*** | 0.026*** | 0.050*** |
| | (0.010) | (0.008) | (0.008) | (0.010) | (0.008) | (0.008) |
| Last Author Collaborator Count (ln) | -0.003 | -0.005 | 0.039** | 0.004 | -0.004 | 0.039** |
| | (0.009) | (0.004) | (0.014) | (0.009) | (0.005) | (0.014) |
| Last Author Prior Team Size (ln) | -0.095*** | -0.065*** | -0.184*** | -0.095*** | -0.065*** | -0.184*** |
| | (0.027) | (0.004) | (0.028) | (0.027) | (0.004) | (0.028) |
| Last Author Prior Team Age (ln) | -0.053*** | -0.046*** | -0.027 | -0.053*** | -0.046*** | -0.027 |
| | (0.009) | (0.007) | (0.018) | (0.009) | (0.008) | (0.017) |
| Last Author Previous Avg. Impact (ln) | 0.104*** | 0.075*** | 0.220*** | 0.104*** | 0.075*** | 0.220*** |
| | (0.028) | (0.004) | (0.021) | (0.028) | (0.004) | (0.021) |
| Last Author English Speaking = True | 0.017*** | 0.010** | 0.055** | 0.017*** | 0.010** | 0.054** |
| | (0.006) | (0.004) | (0.025) | (0.006) | (0.004) | (0.048) |
| Fixed Effects | | | | | | |
| Year-Month | Yes (30 periods) | Yes (30 periods) | Yes (16 periods) | Yes (30 periods) | Yes (30 periods) | Yes (16 periods) |
| Discipline | Yes (19 fields) | Yes (19 fields) | Yes (17 fields) | Yes (19 fields) | Yes (19 fields) | Yes (17 fields) |
| Journal | Yes (29 venues) | Yes (14 venues) | Yes (15 venues) | Yes (29 venues) | Yes (14 venues) | Yes (15 venues) |
| Model Fits | | | | | | |
| Adj. $R^2$ | 0.426 | 0.094 | 0.369 | 0.426 | 0.095 | 0.369 |
| Obs. | 63,551 | 48,968 | 14,583 | 63,551 | 48,968 | 14,583 |

Note: Coefficients are unstandardized estimates. Standard errors are clustered by journal and year-month, and are reported in parentheses. Significance levels: * p < 0.1, ** p < 0.05, *** p < 0.01. The AI usage score is a continuous variable ranging from 0 to 1.

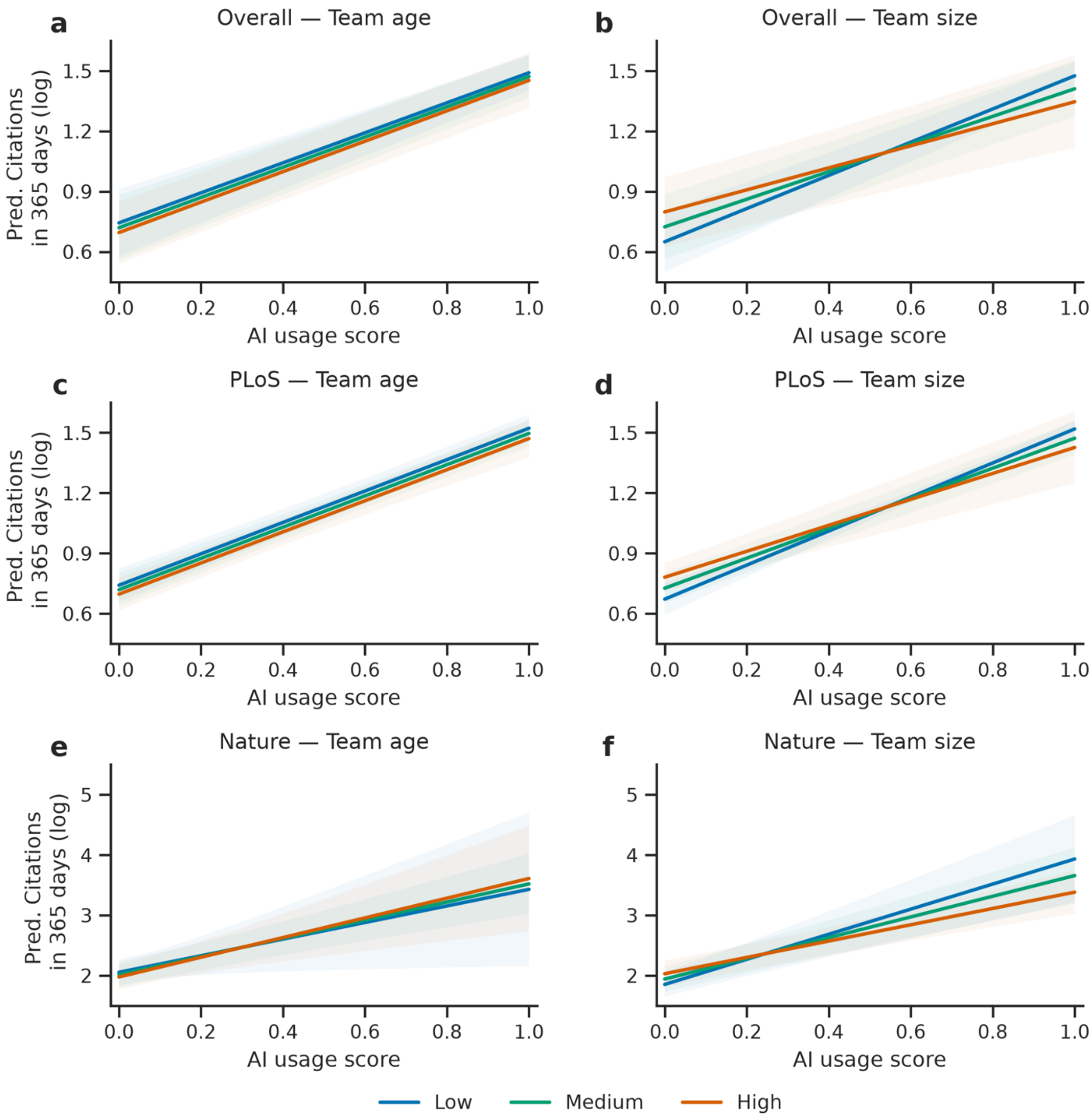

**Figure 7. The estimated interaction of AI usage score on the relationship between team characteristics and short-term scientific visibility.**

Note: Low, Medium and High means the mean of AI usage score minus one standard deviation, mean and the mean plus one standard deviation, respectively. Error bar means 95% CI.

### 4.4 Robustness checks

To assess the sensitivity of our findings to the timing cutoff, classification threshold, and observation window, we conducted several robustness checks.

For the descriptive group-based comparisons, we conducted three alternative robustness checks. First, we shifted the start of the post-ChatGPT period to June 1, 2023, allowing more time for ChatGPT adoption and accounting for the lag between manuscript preparation and publication. We also lowered the classification threshold from the 95th to the 90th percentile of the pre-ChatGPT AI usage score distribution (Supplementary Figures S2, S4, and S6). Second, we retained the original November 30, 2022, publication cutoff but applied a stricter threshold based on the 97.5th percentile of the pre-ChatGPT score distribution (See Supplementary Figure S3, S5, and S7). Third, while keeping the baseline cutoff and 95th percentile threshold, we shortened the citation window from one year to half a year (180 days) to test the sensitivity of our short-term citation measurement (Supplementary Figure S8).

In addition, for regression analyses, we re-estimated the models by restricting to post-June 1, 2023 cohort, which allows a lag between manuscript preparation and publications (Details in the Supplementary Tables S5-S8). Additionally, we shortened the length of citation window from one year to 180 days to investigate if the choices of the length of citation window change the results (See Supplementary Tables S9-10).

In general, all the robustness check results are consisted with the major findings.

## 5 Discussion and conclusion

Using an initial corpus of 144,930 full-text publications from the PLoS family and the Nature Portfolio published between 2020 and 2025, this study examined the associations between AI usage score, publication team characteristics, and citations in one-year citation window. In this study, we consider AI usage score as an observable signal for detecting AI-assisted writing.

We find that higher AI usage scores tend to be associated with smaller and younger research teams. Traditionally, larger research teams can distribute writing, editing, and other communication-related tasks among more members, while researchers with greater academic experience may draw on accumulated knowledge and expertise to prepare and communicate research findings. AI writing tools may provide additional support for these activities, potentially reducing the extent to which researchers rely on larger teams or established academic experience when preparing manuscripts. From this perspective, AI-assisted writing may be particularly relevant to smaller and younger teams that have fewer human resources or less accumulated experience.

However, these findings should not be interpreted as evidence that AI replaces human collaboration or academic expertise. Scientific knowledge production involves much more than writing, including research design, data collection, methodological expertise, and intellectual exchange. The observed relationships may therefore reflect differences in how research teams organize and communicate their work, rather than changes in the underlying organization of scientific inquiry. These possibilities suggest that AI-assisted writing should be understood as one element within a broader process of changing scholarly communication practices.

The second important finding concerns the positive association between AI usage score and the short-term citation of scientific publications. This finding adds a communication-oriented perspective to existing discussions about the implications of GenAI for scientific research. Previous research has largely examined researchers' adoption of AI, their perceptions of its usefulness, and its potential implications for research productivity. Our results suggest that textual signals of AI-assisted writing are also associated with the subsequent short-term visibility of scientific publications.

Several possible mechanisms may help explain this association. AI writing tools can support language refinement, and facilitate the presentation of research findings. These capabilities may help authors communicate scientific contributions more effectively to broader research audiences. AI-assisted writing may also reduce some of the linguistic burdens associated with preparing scientific manuscripts, particularly for researchers working across different linguistic and institutional contexts (Liu et al., 2025). Better communication may, in turn, be associated with greater discoverability, accessibility, and subsequent scholarly engagement. Nevertheless, citation counts should not be equated with research quality. Citation accumulation is also shaped by research topics, disciplinary conventions, journal characteristics, collaboration networks, and the existing visibility of authors (Tahamtan et al., 2016). Publications with greater visibility may differ from other publications in ways that are also associated with textual AI signals.

Furthermore, we examined whether AI usage moderates the relationships between team characteristics and citation visibility. The results reveal a distinction between team academic age and team size. The interaction between AI usage score and team academic age is not statistically significant, suggesting that the association between teams' accumulated academic experience and citation visibility does not vary systematically with AI usage. In contrast, the interaction between AI usage score and team size is consistently negative across all three samples. This indicates that the citation visibility advantage associated with larger teams becomes weaker at higher levels of AI usage. One possible explanation is that AI-assisted writing may help smaller teams undertake writing and communication tasks with fewer human resources, potentially reducing their reliance on the resource and network advantages typically associated with larger teams. These findings extend research on AI in scholarly communication by suggesting that AI usage may be associated not only with citation visibility itself, but also with differences in how team characteristics relate to scientific recognition.

This study has several implications for research on scholarly communication and the organization of scientific knowledge production. Theoretically, this study extends research on GenAI in science beyond its predominantly productivity-oriented focus by introducing a scholarly communication perspective. While existing studies have largely examined how GenAI relates to research productivity, scientific innovation, and researchers' work practices (Liu et al., 2026; Qin et al., 2026), less attention has been paid to how AI-assisted writing relates to the organization, communication, and visibility of scientific knowledge. By examining publication-team composition alongside subsequent citation visibility, this study connects AI-assisted writing to two interconnected dimensions of scholarly communication: the organization of knowledge production and the dissemination and recognition of scientific outputs. In particular, our findings suggest that the association between team size and citation visibility varies with the extent of AI-associated writing. This highlights the importance of considering how emerging communication technologies may reshape the relationship between research team characteristics and the visibility of scientific knowledge.

Practically, the findings are relevant to researchers and research institutions considering how GenAI can be incorporated into scholarly communication. AI writing tools may provide additional support for manuscript preparation, language editing, and the presentation of research findings. Such support may be particularly useful for smaller research teams with limited human resources. However, access to AI tools alone does not guarantee comparable research capabilities or scientific recognition (Channing & Ghosh, 2026). Institutions should therefore consider how researchers with different levels of resources and institutional support can access and appropriately use these technologies.

From a policy perspective, the findings highlight the importance of considering the distributional implications of GenAI in science. Institutional and national policies should move beyond simply encouraging or restricting AI adoption and pay greater attention to the conditions under which researchers use these tools. This includes promoting equitable access to AI-assisted writing resources, providing guidance on responsible use and disclosure (He & Bu, 2026), and maintaining appropriate standards of research integrity and authorship accountability (Chen et al., 2024). Furthermore, research evaluation systems should avoid treating textual signals of AI-assisted writing as direct indicators of research quality or scientific contribution.

This study has several limitations. First, our sample covers only the PLoS family and the Nature Portfolio, limiting the generalizability of the findings. Second, AI-assisted writing is estimated from publication text rather than directly observed, and the observational design does not support causal conclusions. Although we used regression controls and PSM, unobserved confounding and measurement error may remain. Third, we measured publication team characteristics using authorship information, which does not fully capture actual collaboration, expertise diversity, or the division of labor. Finally, we measured short-term citations using a fixed 365-day window. While this ensures comparable citation accumulation periods, it cannot capture long-term scientific influence or research quality. Future studies could combine direct evidence of AI use with more detailed collaboration data and longer-term impact measures.

## Appendix

Please see the attached Appendix for supplementary tables and figures.

## Acknowledgement

The authors used ChatGPT (GPT-6) to assist with language editing and grammatical improvements.

## References

Aksnes, D. W., Langfeldt, L., & Wouters, P. (2019). Citations, citation indicators, and research quality: An overview of basic concepts and theories. *Sage Open*, *9*(1), 2158244019829575.

Andersen, J. P., Degn, L., Fishberg, R., Graversen, E. K., Horbach, S. P., Schmidt, E. K., Schneider, J. W., & Sørensen, M. P. (2025). Generative Artificial Intelligence (GenAI) in the research process–A survey of researchers' practices and perceptions. *Technology in Society*, *81*, 102813.

Ante, L. (2022). The relationship between readability and scientific impact: Evidence from emerging technology discourses. *Journal of Informetrics*, *16*(1), 101252. https://doi.org/https://doi.org/10.1016/j.joi.2022.101252

Arroyo-Machado, W., Ma, J., Chen, T., Johnson, T. P., Islam, S., Michalegko, L., & Welch, E. (2025). Generative AI and academic scientists in US universities: Perception, experience, and adoption intentions. *PloS one*, *20*(8), e0330416.

Bao, T., Zhao, Y., Mao, J., & Zhang, C. (2025). Examining linguistic shifts in academic writing before and after the launch of ChatGPT: a study on preprint papers. *Scientometrics*, *130*(7), 3597-3627.

Ben-Zion, Y., Cohen, E., & Davidovitch, N. (2026). Changes in Manuscript Length, Research Team Size, and International Collaboration in the Post-2022 Period: Evidence from PLOS ONE. *arXiv preprint arXiv:2603.01718*.

Bianchini, S., Müller, M., & Pelletier, P. (2025). Drivers and barriers of AI adoption and use in scientific research. *Technological Forecasting and Social Change*, *220*, 124303. https://doi.org/https://doi.org/10.1016/j.techfore.2025.124303

Binz, M., Alaniz, S., Roskies, A., Aczel, B., Bergstrom, C. T., Allen, C., Schad, D., Wulff, D., West, J. D., Zhang, Q., Shiffrin, R. M., Gershman, S. J., Popov, V., Bender, E. M., Marelli, M., Botvinick, M. M., Akata, Z., & Schulz, E. (2025). How should the advancement of large language models affect the practice of science? *Proceedings of the National Academy of Sciences*, *122*(5), e2401227121. https://doi.org/doi:10.1073/pnas.2401227121

Borgman, C. L. (2007). *Scholarship in the digital age: Information, infrastructure, and the Internet*. The MIT Press.

Caliendo, M., & Kopeinig, S. (2008). Some practical guidance for the implementation of propensity score matching. *Journal of economic surveys*, *22*(1), 31-72.

Channing, G., & Ghosh, A. (2026). AI for scientific discovery is a social problem. *Patterns*, *7*(3), 101497. https://doi.org/https://doi.org/10.1016/j.patter.2026.101497

Chen, Z., Chen, C., Yang, G., He, X., Chi, X., Zeng, Z., & Chen, X. (2024). Research integrity in the era of artificial intelligence: Challenges and responses. *Medicine (Baltimore)*, *103*(27), e38811. https://doi.org/10.1097/md.0000000000038811

Culbert, J. H., Hobert, A., Jahn, N., Haupka, N., Schmidt, M., Donner, P., & Mayr, P. (2025). Reference coverage analysis of OpenAlex compared to Web of Science and Scopus. *Scientometrics*, *130*(4), 2475-2492.

Di Bitetti, M. S., & Ferreras, J. A. (2017). Publish (in English) or perish: The effect on citation rate of using languages other than English in scientific publications. *Ambio*, *46*(1), 121-127.

Ding, W. W., Levin, S. G., Stephan, P. E., & Winkler, A. E. (2010). The impact of information technology on academic scientists' productivity and collaboration patterns. *Management Science*, *56*(9), 1439-1461.

Dorta-González, P., López-Puig, A. J., Dorta-González, M. I., & González-Betancor, S. M. (2024). Generative artificial intelligence usage by researchers at work: Effects of gender, career stage, type of workplace, and perceived barriers. *Telematics and Informatics*, *94*, 102187.

Drubin, D. G., & Kellogg, D. R. (2012). English as the universal language of science: opportunities and challenges. *Molecular biology of the cell*, *23*(8), 1399-1399.

Dwivedi, Y. K., Kshetri, N., Hughes, L., Slade, E. L., Jeyaraj, A., Kar, A. K., Baabdullah, A. M., Koohang, A., Raghavan, V., & Ahuja, M. (2023). Opinion Paper:"So what if ChatGPT wrote it?" Multidisciplinary perspectives on opportunities, challenges and implications of generative conversational AI for research, practice and policy. *International journal of information management*, *71*, 102642.

Gross, D. P., Shvadron, D., & Zhang, H. (2026). *How heavily are PhD students offloading their scientific writing to AI? Evidence from US STEM PhD dissertations*.

Harries, R., Lawson, C., & Shapira, P. (2025). Generative AI in science: Applications, challenges, and emerging questions. *arXiv preprint arXiv:2507.08310*.

He, Y., & Bu, Y. (2026). Academic journals' AI policies fail to curb the surge in AI-assisted academic writing. *Proceedings of the National Academy of Sciences*, *123*(9), e2526734123.

Hey, T. (2023). Artificial intelligence for science and engineering: A priority for public investment in research and development.

Hosseinioun, M., Uzzi, B., & Fosse, H. B. (2026). Artificial Intelligence in Science: Returns, Reallocation, and Reorganization. *arXiv preprint arXiv:2603.27956*.

Hsiehchen, D., Espinoza, M., & Hsieh, A. (2015). Multinational teams and diseconomies of scale in collaborative research. *Science Advances*, *1*(8), e1500211.

Ibrahim, S. K. M., & Mahmoud, Z. A. Z. (2026). Generative AI in academic writing: a comparison of human-authored and ChatGPT-generated research article titles. *Humanities and Social Sciences Communications*, *13*(1), 394. https://doi.org/10.1057/s41599-026-06956-z

Koenker, R., & Hallock, K. F. (2001). Quantile regression. *Journal of economic perspectives*, *15*(4), 143-156.

Kosteas, V. D. (2018). Predicting long-run citation counts for articles in top economics journals. *Scientometrics*, *115*(3), 1395-1412.

Kusumegi, K., Yang, X., Ginsparg, P., de Vaan, M., Stuart, T., & Yin, Y. (2025). Scientific production in the era of large language models. *Science*, *390*(6779), 1240-1243. https://doi.org/doi:10.1126/science.adw3000

Larivière, V., Gingras, Y., Sugimoto, C. R., & Tsou, A. (2015). Team size matters: Collaboration and scientific impact since 1900. *Journal of the Association for Information Science and Technology*, *66*(7), 1323-1332. https://doi.org/https://doi.org/10.1002/asi.23266

Lee, Y.-N., Walsh, J. P., & Wang, J. (2015). Creativity in scientific teams: Unpacking novelty and impact. *Research Policy*, *44*(3), 684-697. https://doi.org/https://doi.org/10.1016/j.respol.2014.10.007

Lei, L., & Yan, S. (2016). Readability and citations in information science: Evidence from abstracts and articles of four journals (2003–2012). *Scientometrics*, *108*(3), 1155-1169.

Liang, W., Izzo, Z., Zhang, Y., Lepp, H., Cao, H., Zhao, X., Chen, L., Ye, H., Liu, S., & Huang, Z. (2024). Monitoring ai-modified content at scale: A case study on the impact of chatgpt on ai conference peer reviews. *arXiv preprint arXiv:2403.07183*.

Liang, W., Zhang, Y., Wu, Z., Lepp, H., Ji, W., Zhao, X., Cao, H., Liu, S., He, S., & Huang, Z. (2025). Quantifying large language model usage in scientific papers. *Nature Human Behaviour*, *9*(12), 2599-2609.

Liu, J., He, Y., Zheng, Z., Bu, Y., & Ni, C. (2025). AI-assisted writing is growing fastest among non-English-speaking and less established scientists. *arXiv preprint arXiv:2511.15872*.

Liu, M., Bu, Y., Chen, C., Xu, J., Li, D., Leng, Y., Freeman, R. B., Meyer, E. T., Yoon, W., & Sung, M. (2022). Pandemics are catalysts of scientific novelty: Evidence from COVID-19. *Journal of the Association for Information Science and Technology*, *73*(8), 1065-1078.

Liu, M., & Hu, X. (2022). Movers' advantages: The effect of mobility on scientists' productivity and collaboration. *Journal of Informetrics*, *16*(3), 101311.

Liu, M., Jaiswal, A., Bu, Y., Min, C., Yang, S., Liu, Z., Acuña, D., & Ding, Y. (2022). Team formation and team impact: The balance between team freshness and repeat collaboration. *Journal of Informetrics*, *16*(4), 101337. https://doi.org/https://doi.org/10.1016/j.joi.2022.101337

Liu, M., Xie, Z., Yang, A. J., Yu, C., Xu, J., Ding, Y., & Bu, Y. (2024). The prominent and heterogeneous gender disparities in scientific novelty: Evidence from biomedical doctoral theses. *Information Processing & Management*, *61*(4), 103743.

Liu, M., Yang, S., Bu, Y., & Zhang, N. (2023). Female early-career scientists have conducted less interdisciplinary research in the past six decades: evidence from doctoral theses. *Humanities and Social Sciences Communications*, *10*(1), 918. https://doi.org/10.1057/s41599-023-02392-5

Liu, Y., Xie, Y., Shen, X., & Wu, D. (2026). Artificial intelligence use and scientific innovation. *Journal of the Association for Information Science and Technology*, *77*(5), 682-698. https://doi.org/https://doi.org/10.1002/asi.70043

Ma, R., Dedema, M., & Cox, A. (2026). A dancing bear, a colleague, or a sharpened toolbox? The cautious adoption of generative artificial intelligence technologies in digital humanities research. *Journal of the Association for Information Science and Technology*, *77*(6), 812-830.

Marino Fages, D. (2020). Write better, publish better. *Scientometrics*, *122*(3), 1671-1681.

Matsui, A. (2026). Return of the solo author: The changing division of labor in science in the age of generative A. *arXiv preprint arXiv:2607.10780*.
Mayer, T., & Zignago, S. (2011). Notes on CEPII's distances measures: The GeoDist database.
Nasra, M. A., & Oliver, A. L. (2025). Human and social capital and ethnically diverse founding teams in high-tech industries. *The Journal of Technology Transfer*, *50*(3), 1199-1230.
Nicholas, D., Swigon, M., Clark, D., Abrizah, A., Revez, J., Herman, E., Rodríguez Bravo, B., Xu, J., & Watkinson, A. (2024). The impact of generative AI on the scholarly communications of early career researchers: An international, multi-disciplinary study. *Learned Publishing*, *37*(4), e1628.
Nielsen, M. W., Alegria, S., Börjeson, L., Etzkowitz, H., Falk-Krzesinski, H. J., Joshi, A., Leahey, E., Smith-Doerr, L., Woolley, A. W., & Schiebinger, L. (2017). Gender diversity leads to better science. *Proceedings of the National Academy of Sciences*, *114*(8), 1740-1742.
Qian, Y., Wen, Z., Furnas, A. C., Bai, Y., Shao, E., & Wang, D. (2026). The rise of large language models and the direction and impact of US federal research funding. *Proceedings of the National Academy of Sciences*, *123*(33), e2601439123.
Qin, C., Pee, L., Chen, X., & Wang, X. (2026). Emerging research information practices in the era of generative artificial intelligence. *Journal of the Association for Information Science and Technology*.
Robinson-Garcia, N., Costas, R., Sugimoto, C. R., Larivière, V., & Nane, G. F. (2020). Task specialization across research careers. *Elife*, *9*, e60586.
Ryba, R., Doubleday, Z. A., & Connell, S. D. (2019). How can we boost the impact of publications? Try better writing. *Proceedings of the National Academy of Sciences*, *116*(2), 341-343.
Stavrova, O., Kleinberg, B., Evans, A. M., & Ivanović, M. (2025). Scientific publications that use promotional language in the abstract receive more citations and public attention. *Communications Psychology*, *3*(1), 118.
Tahamtan, I., Safipour Afshar, A., & Ahamdzadeh, K. (2016). Factors affecting number of citations: a comprehensive review of the literature. *Scientometrics*, *107*(3), 1195-1225.
Taşkın, Z. (2026). Sustaining the "frozen footprints" of scholarly communication through open citations: An Annual Review of Information Science and Technology (ARIST) paper. *Journal of the Association for Information Science and Technology*, *77*(1), 23-39.
Thelwall, M., Kousha, K., Stuart, E., Makita, M., Abdoli, M., Wilson, P., & Levitt, J. (2023). In which fields are citations indicators of research quality? *Journal of the Association for Information Science and Technology*, *74*(8), 941-953.
Thelwall, M., & Maflahi, N. (2022). Research coauthorship 1900–2020: Continuous, universal, and ongoing expansion. *Quantitative Science Studies*, *3*(2), 331-344. https://doi.org/10.1162/qss_a_00188
Tomaszewski, R. (2023). Visibility, impact, and applications of bibliometric software tools through citation analysis. *Scientometrics*, *128*(7), 4007-4028.
Wang, H., Fu, T., Du, Y., Gao, W., Huang, K., Liu, Z., Chandak, P., Liu, S., Van Katwyk, P., Deac, A., Anandkumar, A., Bergen, K., Gomes, C. P., Ho, S., Kohli, P., Lasenby, J., Leskovec, J., Liu, T.-Y., Manrai, A.,…Zitnik, M. (2023). Scientific discovery in the age of artificial intelligence. *Nature*, *620*(7972), 47-60. https://doi.org/10.1038/s41586-023-06221-2
Wu, L., & Vasilescu, B. (2026). AI raises the productivity bar. *Science*, *391*(6787), 763-764. https://doi.org/doi:10.1126/science.aef5239
Wu, L., Wang, D., & Evans, J. A. (2019). Large teams develop and small teams disrupt science and technology. *Nature*, *566*(7744), 378-382.
Wuchty, S., Jones, B. F., & Uzzi, B. (2007). The Increasing Dominance of Teams in Production of Knowledge. *Science*, *316*(5827), 1036-1039. https://doi.org/doi:10.1126/science.1136099
Xu, F., Wu, L., & Evans, J. (2022). Flat teams drive scientific innovation. *Proceedings of the National Academy of Sciences*, *119*(23), e2200927119.
Xu, H., Bu, Y., Liu, M., Zhang, C., Sun, M., Zhang, Y., Meyer, E., Salas, E., & Ding, Y. (2022). Team power dynamics and team impact: New perspectives on scientific collaboration using career age as a proxy for team power. *Journal of the Association for Information Science and Technology*, *73*(10), 1489-1505.

Yang, A. J., Xu, H., Ding, Y., & Liu, M. (2024). Unveiling the dynamics of team age structure and its impact on scientific innovation. *Scientometrics*, *129*(10), 6127-6148.

Yang, Y., Tian, T. Y., Woodruff, T. K., Jones, B. F., & Uzzi, B. (2022). Gender-diverse teams produce more novel and higher-impact scientific ideas. *Proceedings of the National Academy of Sciences*, *119*(36), e2200841119.

Zhao, N., & Li, L. (2026). The diverging effect of prestige and experience on the use of artificial intelligence knowledge. *Journal of the Association for Information Science and Technology*, *77*(4), 580-595. https://doi.org/https://doi.org/10.1002/asi.70068

Zheng, X., Hong, X., Liu, J., & Ni, C. (2026). Scientific exploration, collaboration and labor division in the large language model era. *arXiv preprint arXiv:2607.20923*.